# The formation of disks in massive spiral galaxies

François Hammer[a] with Mathieu Puech[a] & Hector Flores[a]

[a] *Laboratoire Galaxies Etoiles Physique et Instrumentation, Observatoire de Paris, UMR8111 du CNRS, 5 place Jules Janssen, 92195 Meudon France*

**Abstract.** The flatness of the rotation curve inside spiral galaxies is interpreted as the imprint of a halo of invisible matter. Using the deepest observations of distant galaxies, we have investigated how large disks could have been formed. Observations include spatially resolved kinematics, detailed morphologies and photometry from UV to mid-IR. Six Giga-years ago, half of the present-day spirals had anomalous kinematics and morphologies that considerably affect the scatter of the Tully Fisher relation. All anomalous galaxies can be modelled through gas-rich, major mergers that lead to a rebuilt of a new disk. The spiral-rebuilding scenario is proposed as a new channel to form large disks in present-day spirals and it accounts for all the observed evolutions since the last 6 Giga-years. A large fraction of the star formation is linked to merging events during their whole durations.

## INTRODUCTION

A considerable fraction, 72%, of massive galaxies is made of spiral galaxies, including the Milky Way and M31 in the Local Group. By massive galaxies, I arbitrarily consider those with stellar masses larger than $10^{10}M_{sun}$, the Milky Way being five times more massive than this value. Since the early 20$^{th}$ century it is well known that the baryonic matter is not massive enough to equilibrate the observed rotation of stars in the giant disks, requiring the presence of a massive unseen component [1]. Besides the important question of the nature of the dark matter, astrophysicists have debated a lot the following question: how large disks formed in massive spirals? Disks are supported by their angular momentum that may be acquired by early interactions in the framework of the tidal torque theory [2,3]. In this theory, galactic disks are then assumed to evolve without subsequent major mergers, in a secular way.

However this theory faces with at least two major problems. First, galaxy simulations demonstrate that such disks can be easily destroyed by collisions of relatively big satellites[1] [4], and such collisions might be too frequent to let disks survive; Second, the disks produced by simulations in such a way are too small or have a too small angular momentum when compared to the observed ones, and this is so-called the "angular momentum catastrophe". Besides this, an early and gradual formation of the disk of the Milky Way is well supported by observations.

## THE MILKY WAY IS AN EXCEPTIONAL LOCAL SPIRAL

Observations have considerably improved since the early 70s and it is now possible to analyze with much more details the properties of progenitors of present-day galaxies by observing distant galaxies. Moreover our knowledge of the two massive spirals in our neigbourhood, the Milky Way and M31, has considerably improved. It is now well established that the Milky Way experienced very few minor mergers and no major merger during the last 10-11 Gyrs

[1] In the following I will assume that major mergers are those which destroy the disk of the main interloper, which correspond to various mass ratio depending on the orbital parameters; generally major mergers are assumed to have mass ratio larger than 1:5, 1:4 or 1:3.

[5,6]. The old stellar content of the thick disk let possible a merger origin at that early epoch. The Milky Way is presently absorbing the Sagittarius dwarf though this is a very tiny event given the fact that the Sagittarius mass is less than 1% of the Milky Way mass [7]. If the Magellanic Clouds are truly bound to the Milky Way, a collision is expected within the next Giga-year, which may have more impact to the Milky Way disk since they carry on a quite significant amount of orbital angular momentum. Milky Way past history strongly contrasts with that of M31 (Andromeda). Numerous works have revealed the very tumultuous history of M31 [8, 9, 10]. Its outskirts is surrounded by gigantic structures at low surface brightness, including the Giant stream, the extended thick disk, rings etc.. Several attempts have been successfully made to relate some of these structures to interactions with smaller neighbouring galaxies (e.g. M33). However we are still lacking of a full model of M31 and its outskirts, possibly because its past history was too rich in merger events. It is still uncertain whether M31 experienced only several minor interactions, or also a major merger.

Observations of galaxies 'outskirts at similar depths are in progress, and superb structures have been identified surrounding NGC 4013, NGC 5907, M 63 and M 81 [11, 12]. They are all reminiscences of past mergers events. It has to be established whether or not most spiral galaxies are showing similar richness of past merger events in their halo. It would be also interesting to simulate how would appear the Milky Way and the Sagittarius stream from an external observer. There is an important clue that many spirals' haloes contain more relics of past merger events than does the Milky Way. In fact the discovery that RGB stars in the halo in very nearby spirals [13] are redder than those of the Milky Way implies that they have been metal enriched by a way or another. The same applies to the important population of local spirals from the SDSS as it has been shown by [14] after stacking more than one thousand spiral halo emission. Figure 1 displays the corresponding metal abundance of stars in the inner halo of various galaxies, including the Milky Way and M31. In all massive galaxies with $V_{flat}$ > 150 km/s, halo stars have metal abundances much larger than those of the Milky Way. Most of these spiral galaxies have been observed by [13] and are very nearby. The agreement with the average value from SDSS brings a lot of support to the idea that most spirals have an inner halo much more enriched than that of the Milky Way. NGC 891 has been observed by [17] because it shows a lot of similarities with the Milky Way, although its pristine halo is not discovered yet. Examination of the very outskirts of M31 has revealed [16] the existence of a pristine halo at very large distances, up to 150 kpc. The M31 halo, as well as its globular cluster system, is similar to the sum of a pristine halo, or the Milky Way halo, with numerous metal enriched stars, probably due to merger residuals. Conversely, the Milky Way halo and globular cluster system may require the absence of previous merger of satellites with mass larger or equal to $10^9\,M_{sun}$ [18].

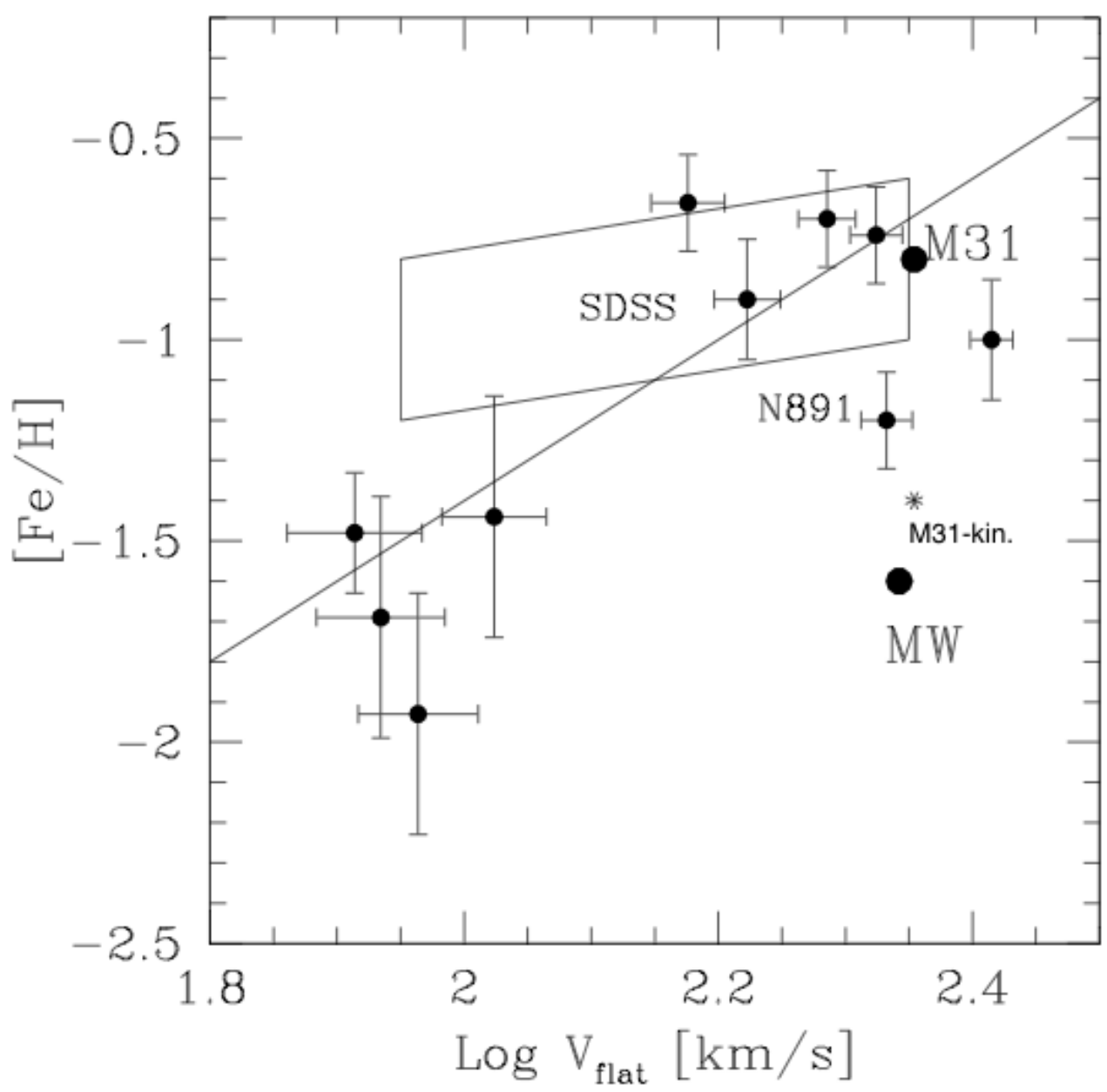

**FIGURE 1.** Iron abundances of RGB stars estimated in the outskirts (2-10kpc from the minor axis) of edge-spirals galaxies from [13] plotted against log $V_{flat}$. Large points represent the values for the Milky Way and M31. The solid lines delineate the range of values from [14], after stacking 1047 edge-on SDSS galaxies. This Figure is similar to Figure 6 of [15], to which we have added the value for the pristine halo of M31 that has been recovered at 60-150 kpc ([16]). NGC 891 measurements from [17] are also added, as this galaxy shows many similarities with the Milky Way. The vertical error bars have been taken from [13].

Thus the Milky Way outskirts appear to be quite exceptional. It is useful to compare the fundamental parameters of its disk to those of other galaxies. The main difficulty is coming from the fact that we are lying in the Milky Way, and then this is not an easy task. By chance, very detailed models of the light distribution of the Galaxy were necessary to remove at best its signal to recover the CMB emission. Hipparcos also provided very useful data to model in detail the Galaxy [18]. Using these data, [15] have shown that the Galactic disk radius, $R_d$= 2.3 +/- 0.6 kpc, is quite small especially when compared to that of M31 ($R_d$= 5.8 +/- 0.6 kpc). The whole emission of both the Milky Way and M31 in K-band have been well recovered by Spitzer, and provides $M_K$=-24.02 and -24.51, for the Milky Way and M31, respectively. The difference between the two values indicates that the stellar mass of M31 is twice that of the Milky Way, after accounting for their respective stellar mass to K-band luminosity ratios. Even if the Milky Way is approximately twice gas rich than M31, the baryonic mass ratio is still 2, because the gas content is rather marginal in both galaxies. Interestingly the two galaxies have the same circular, or flat velocities, implying a different location in the fundamental volume delineated by mass, radius and velocity. On the basis of a very detailed study of the local fundamental relations (mass-velocity or Tully Fisher, radius-velocity) for local spirals, [15] showed that M31 is quite a typical spiral, while the Milky Way is surprisingly exceptional[2], being offset by $1\sigma$ in both relationships. To summarize, only 7% of local spirals have similar properties than the Milky Way.

Some interesting lessons can be learned. The Milky Way is lacking approximately half of its stellar mass, radius and angular momentum when compared to an average spiral, or to M31. Its halo and globular system is pristine, almost free of merger residuals in contrast to the vast majority of other galaxies that systematically show redder halo stars, or more metal-enriched RGB stars. This is quite suggestive that the vast majority of spirals experienced a quite rich merger history, as M31 certainly did, contrasting to the well-established quiescent history of the Milky Way. This would explain why the Milky Way is lacking of angular momentum when compared to other galaxies. Indeed by accounting for the additional effect of the orbital angular momentum provided by major mergers, [19] showed that the angular momentum catastrophe could be solved.

Future observations of halo stars in a larger volume will bring a lot of new constraints to the above, as well as new measurements of the detailed structure of the Galaxy and of M31. Within our present knowledge the Galaxy is not an ordinary spiral, and understanding the formation of large disks in massive spirals requires observing in great details other spirals. Consequently, a gradual elaboration of large disks might not be the rule for most disks, and we may have to identify different channels for their formation. Another method to investigate the disk formation is to verify directly whether or not many spirals have experienced merger events. The present generation of very large telescopes allows us to study in detail galaxies at large look-back times, which have to be in a statistical sense the progenitors of the present-day spirals.

## WHAT IS THE PAST HISTORY OF PRESENT-DAY SPIRALS?

There have been many studies of distant galaxies, and reviewing them is out of the scope of this paper. I would simply mention the first significant survey of distant galaxies up to z=1, the Canada France Redshift Survey, that evidenced for the first time the strong decrease of the star formation density since the last 8 Giga-years [20]. This has been confirmed by infrared photometry [21], using ISO and more recently Spitzer. Integration of star formation density over look-back time can be compared to evolution of the stellar mass density, which is obtained from near-IR photometry [22, 23]. Both approaches give the same result, which could be considered as robust: 50% of the present-day stellar mass has been formed at $z < 1$ or approximately since the last 8 Gyrs. Analyses by [24, 25] of representative samples selected in near-IR and observed in mid-IR have revealed that the star formation below z=1 is dominated by luminous IR galaxies (LIRGs) with star formation rates from 19 to 190 $M_{sun}$/year. Because the fraction of red E/S0 does not change from z=1 to z=0, the stars formed during the last 8 Gyrs are mostly assembled in spiral galaxies [24, 26], and this is confirmed by analyses of LIRG morphologies that range from peculiar or merger, to massive spirals.

[2] I have to mention here that both John Kormendy and Joe Silk told me that Allan Sandage firstly mentioned the hypothesis of an exceptional Milky Way.

## A Strategy to Observe Distant Galaxies and to Relate Them to Local Galaxies

*Observational Requirements*

A remarkable suite of telescopes and instruments is now available to study the distant Universe. As a European scientist I may focus -in a quite biased manner- on the Very Large Telescopes operated by ESO. These 4 telescopes offer a suite of 12 instruments allowing us to cover very large wavelength, spectral and spatial resolution ranges and give us the possibility to observe several sources at the same time. FLAMES/GIRAFFE is successfully operating since July 2002 at the UT2 of the VLT. It is still the only instrument offering to observe, simultaneously up to 15 galaxies using integral field units (IFUs). Using such an instrument, our team has been able to gather the largest existing sample of velocity fields of distant galaxies (see Figure 2).

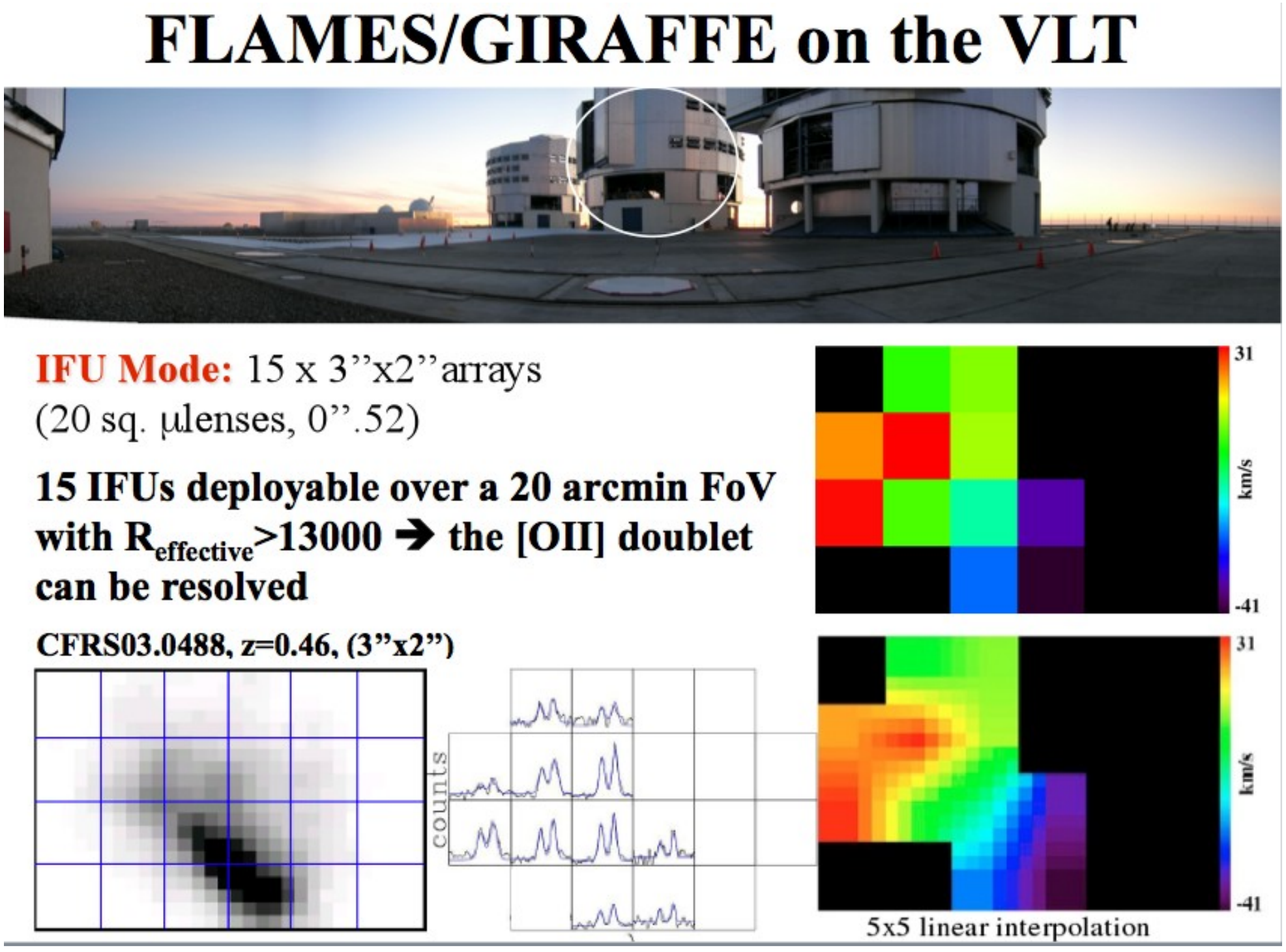

**FIGURE 2.** A sketchy representation of FLAMES/GIRAFFE and its IFU mode is represented. Each galaxy is dissected in up to 20 small apertures or micro-lenses, each of them providing a spectrum at intermediate resolution. In the illustration the OII doublet is well resolved, and each line can be used to trace the velocity and the dispersion of an individual galaxy region. In the right panels the corresponding distribution of the velocity field is shown, that has been interpolated for practical reason in the bottom panel.

However kinematics is not all and other important information need to be collected by other facilities. The best strategy is to perform kinematic surveys within existing cosmological fields that have been surveyed by HST, Spitzer and other telescopes. Because distant galaxies may present various evolutionary stages, it is mandatory to compare kinematical maps to morphological details. For example, the dynamical axis has to be compared to the disk optical axis: if they are severely misaligned, the galaxy is likely not at equilibrium, or not relaxed. The Hubble Space Telescope equipped with the Advanced Survey Camera is able to provide details of few hundred of parsecs at $z < 1$. However it is only a 2.5 meters telescope and depth is an important issue for distant galaxies whose emissions are severely affected by cosmological dimming. For example, the optical radius (3.2 times the disk scale length) of a Milky Way like galaxy at z=0.5 (z=1.3) requires 3 (100) hours of HST/ACS observations to be properly recovered, and this is without accounting for extinction effects. To be predictive enough, studies of kinematics at $z > 1$ would have to be undertaken in extremely deep fields, and up to now, there is only the small Ultra Deep Field.

In the following I will focus on the comparison between z ~ 0.65 and local galaxies. Recall that z= 0.65 represents a 6 Gyrs look-back time, which is not negligible. Within this redshift range, we may indeed compare galaxies in a quite homogeneous manner: most quantities can be evaluated in a similar way, independently of the redshift[3]. Figure 2 illustrates that kinematics of distant galaxies implies to dissect each of them in several parts. Given the faintness of these galaxies, large amounts of time are needed to reach useful S/N. Using GIRAFFE at VLT, [27, 28] showed that from 8 to 24 hours of exposure time are requested to recover the kinematics of almost all emission line galaxies with S/N ≥ 4 in at least 5 pixels or micro lenses. A special care is needed to select samples of galaxies for which we need to examine together their velocity fields, their morphologies and their masses.

### *Representative Samples of Local and Distant Galaxies*

First, we need to determine the simplest method in selecting galaxies. At first glance, the absolute luminosity in near-IR is a relatively good proxy of the stellar mass. It indeed limits the evolutionary effects of the $M_{stellar}/L$ that would be severe in bluer broad bands. By applying $M_J < -20.3$ to both samples of local and z=0.65 galaxies, [26, see their Appendix A] demonstrated that they essentially select galaxies within the same baryonic mass range, because the evolutionary effect of $M_{stellar}/L_J$ (~ 0.15 dex) is almost exactly compensated by the gas fraction increase with redshift (~ -0.16 dex). Second, we need to ensure that our selection is representative of the luminosity function of galaxies at each redshift. This has been verified by [26, 28] after comparison with various luminosity functions derived from large surveys. Third, we have to account for more subtle effects linked with observational techniques. For example, derivation of kinematics for $M_J < -20.3$ distant galaxies is only feasible for sources with $W_0$(OII) ≥ 15A [27, 28]. We should complement the sample of emission line galaxies by a sample of "quiescent" galaxies with $W_0$(OII) < 15A [26]. The only residual difficulty would be the absence of kinematic information for the distant, quiescent galaxies, out of reach of 8 meters telescopes. This is not a stringent difficulty because 60% of z=0.65 galaxies have $W_0$(OII) ≥ 15A [29]. Fourth, we have to ensure that the resulting selections of galaxies are not affected by cosmological variance. This has been done by [27, 28] by selecting 63 galaxies in 4 different fields of view, including the GOODS-S, the CFRS 03 and 22h and the HDF-S. However some results discussed in the following are based on the sole GOODS-S field (33 galaxies). Besides being the most surveyed field at high Galactic latitude, it shows two large-scale structures at z= 0.668 and z=0.735, which may affect the luminosity density evolution between z=0.25-0.5 and z=0.5-0.75 by up to 0.4-0.5 dex [30]. This should be kept in mind during the discussion of the results based on the sole GOODS-S field.

## The Strong Evolution of Massive Galaxies at z=0.65: Morphologies, Kinematics, Star Formation and Gas & Metal Fraction

### *Morphological Changes of Massive Galaxies since the last 6 Giga-years*

The HST/ACS provides details of 200 pc in z=0.65 galaxies, whose z-band observations can be directly compared to r-band of SDSS galaxies at z=0. Inspections of the morphologies evidence that galaxies have strongly evolved. Peculiar galaxies represent half of z=0.65 galaxies [31, 32, 26], while the fraction of early type galaxies is similar to that at z=0 (see Figure 3). It means that more than half of the progenitors of present-day spirals are peculiar galaxies at z=0.65 [26, 32]. Here I need to make two important statements. First, morphologies should be examined in detail, either from examination by a well-experimented observer [32] or by using a complete and full analysis of the light distribution in several colours (bulge-disk decomposition using GALFIT, see [33]) and of a semi-automatic "decision tree" (see [26] and their Figure 4). Galaxies are too complex systems to have their morphologies classified by a set of only 2 parameters such as compactness or asymmetry: those classifications often overestimate the number of spiral morphologies [34]. Second, it should be pointed out that peculiarities in distant galaxies are generally observed at rest-frame UV to near-IR, meaning that all the galaxy light distribution is affected[4]. Some peculiarities could be due to complex networks of star forming regions as it could be observed in local irregular dwarves. The large variety of peculiar morphologies at such high surface brightness (see Figure 3) could be also attributed to mergers [34]. Solving out this important question requires investigating the internal motions in distant galaxies.

[3] K-correction effects can be simply ignored when observing z=0.65 galaxies in filters with central wavelengths that are 1.65 times that of filters used to observe z=0 galaxies. For spectral energy distribution and star-formation rates this is feasible using Spitzer IRAC & MIPS as well as HST/ACS broad-band filters that could be compared to SDSS AND 2MASS filters (see [26]). Morphological analyses of SDSS can be easily compared to those of distant galaxies in the GOODS-S field.

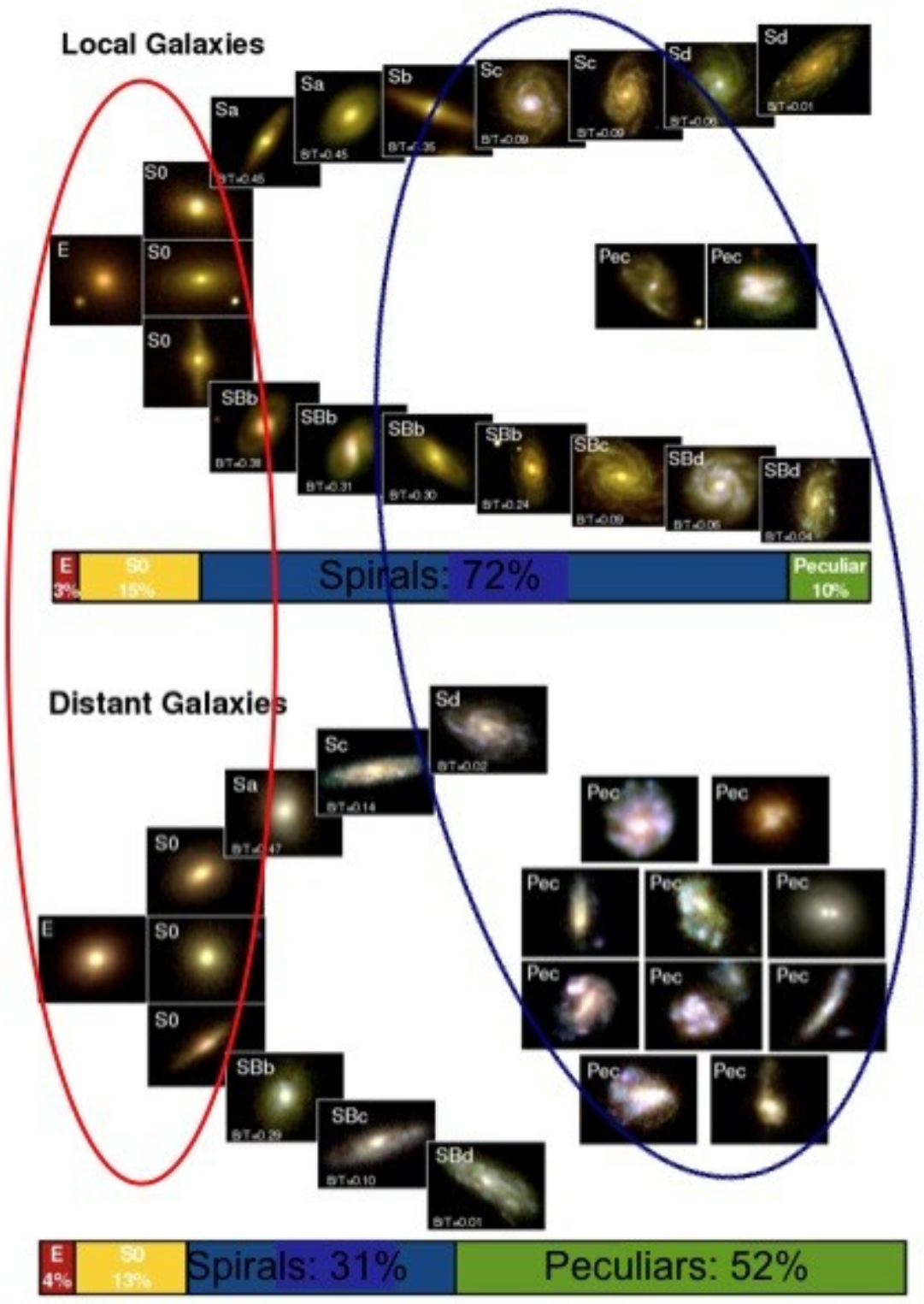

**FIGURE 3.** Morphologies of galaxies at z=0 (top) and z=0.65 (bottom) from [26]. Each stamps represents 5% of the $M_J \leq -20.3$ galaxy population and are 3-colours composite of u, g and r or of v, i and z bands, respectively. The fraction of regular spirals has significantly increased since 6 Gyrs, while for early-type (E/S0) it appears unchanged. At least half of the present-day spirals have a progenitor which is peculiar at z=0.65.

*Strong Evolution of the Kinematics since the last 6 Giga-years*

Kinematical study of distant galaxies is a recent field of investigations. At present all studies are limited by the spatial resolution, i.e. to gas motions at scales from 1 to 3 kilo-parsecs. However, such motions are crucial to disentangle disk rotation from other mechanisms affecting galaxy dynamics at large scales (Figure 4). To reach higher spatial resolution requires combining both adaptive optics techniques and extremely large exposure time, because of the apparent faintness of distant galaxies. GIRAFFE observations are limited by the pixel size (0.52 arc second or 3.5 kpc at z=0.65), although the *precision* in identifying the location of kinematical features is much better than that. This comes from simulations of GIRAFFE observations using redshifted local templates degraded down to the GIRAFFE IFU resolution, and assuming a similar S/N that that from observations of distant galaxies[5]. It shows that kinematic features are stable within variations in positions of one half of an IFU pixel, meaning that the *precision* in recovering their location is close to 0.25 arcsec [36]. This sub-spatial pixel *precision* largely results from the high spectral resolution of GIRAFFE, which provides us with a very accurate kinematic measurement along the spectral axis. Finally we have learnt from our experience with GIRAFFE that the accuracy in comparing galaxy kinematics with morphology is generally dominated by the relative astrometry (0.23 arcsec). The half pixel accuracy has been illustrated in the observation of a satellite (1:18) in falling into a z=0.7 galaxy, responsible of the offset of the dispersion peak [38].

[4] It has been claimed by [35] that distant LIRGs have regular, disk morphologies on the basis of an adaptive optic survey conducted at the Keck telescope. However the exposure time was too limited to allow an observation beyond the core of individual galaxies. An examination of much deeper HST/ACS observations reveals their peculiar morphologies with rings, tidal tails, irregularities and even mergers, in rest-frame V-band.

[5] A generalization of this simulator is now used to prepare several phases A of E-ELT instruments, especially the link between spatial resolution provided by different adaptive optics techniques and spatially resolved kinematics (see [37]).

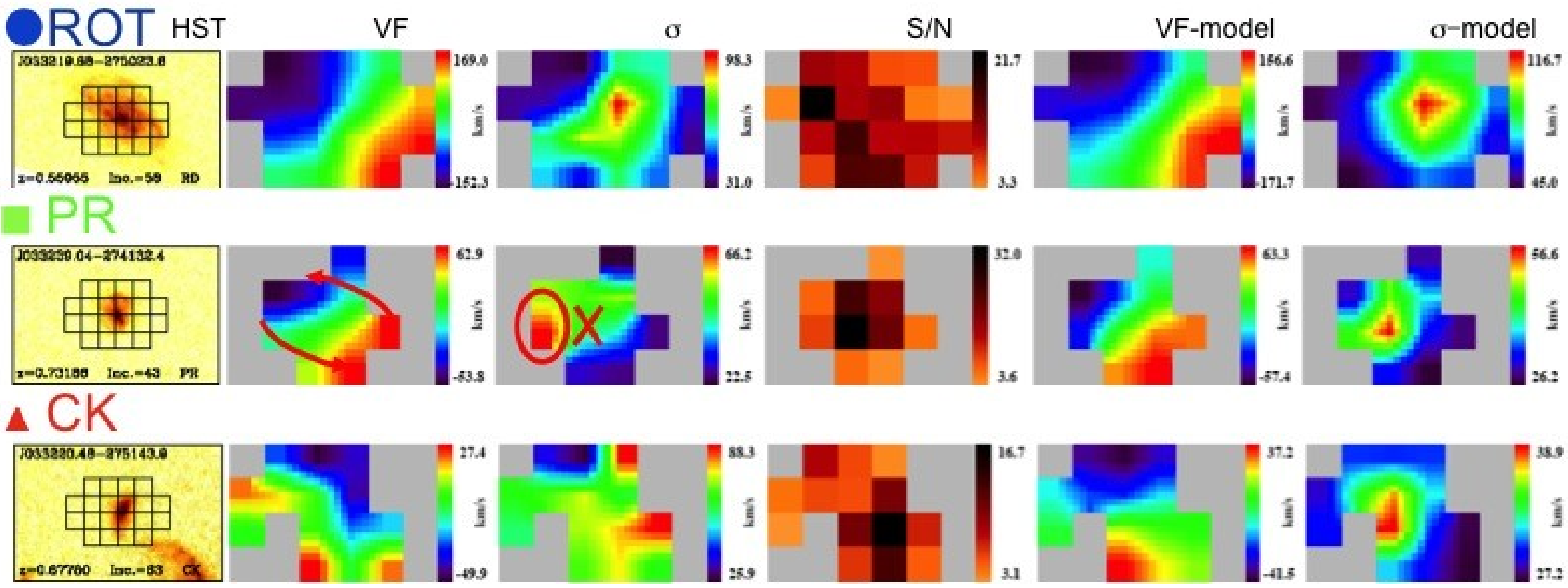

**FIGURE 4.** It illustrates three different examples (or class) of kinematical properties of distant galaxies. The upper row shows a rotating disk showing a dynamical axis aligned with the optical axis. Because of the low spatial resolution most of the gradient between the two extreme velocities provides a dispersion peak (see 3rd and 6th panels from the left) just at the centre of the velocity field and of the galaxy, i.e. the centre of rotation. Velocity models assume the inclination from the optical image and a velocity gradient from the observations. The middle row shows a galaxy with an almost aligned dynamical axis, thus indicative of a rotation. However the dispersion peak falls far from the centre of rotation, indeed far from any optical emission of the galaxy. This is indicative of a strong disturbance of the gaseous emission, probably due to shocks in gas flows, which perturbs the rotation (called perturbed rotation, PR). In the bottom row, the kinematics is complex (called complex kinematics, CK) from both velocity field and dispersion maps. The galaxy is likely after its first passage during its merger with a more massive spiral [39].

Among the 63 z=0.4-0.75 galaxies observed by GIRAFFE, most of them show anomalous kinematics. Accounting for a representative sample of $M_J \leq -20.3$ galaxies (in emission and in absorption) there is only 19% of rotating disks, 15% of perturbed rotators and 26% of complex kinematics, the resting 40% being made of "quiescent" galaxies that could not been observed due to the faintness of their emission lines. The latest galaxies have been examined by [26] and can be further subdivided into 18% of E/S0, 12% of spirals and 10% of peculiar galaxies. Taking into account the non-evolving fraction of E/S0 let us with 72% of present-day spirals and only 31% of rotating disks at z=0.65. Then at least half of the present-day spirals progenitors at z=0.65 have anomalous kinematics. This implies a strong evolution of spiral kinematics[6] with amplitude similar to what is found from morphological analyses.

*Combining Morphological and Kinematical Analyses*

A thorough comparison of morphology and kinematics of 52 distant galaxies has been made by [31]. They find that 92% of galaxies with anomalous kinematics (PR or CK) have peculiar morphologies, and that 80% of rotating galaxies have spiral morphologies. This result stands for morphological classifications assuming a strict similarity in classifying local and distant galaxies [26, 32], but it does not for automatic or semi-automatic methods based on asymmetry and compactness. Nevertheless it has profound implications, because kinematics is sampling the ionised gas ([OII] emission lines) while galaxies morphologies are due to bright stars emitting at all optical wavelengths. Then a common mechanism is responsible of both modes of disturbances in half of the present-day spiral progenitors. In fact the combination of detailed morphologies with HST/ACS with large-scale kinematics from VLT/GIRAFFE is well appropriate. For example, most distant galaxies show numerous blue or dusty regions due to recent star formation (SF). The characteristic SF time (~ 100Myrs) is similar to the characteristic dynamical time to affect the galaxy large-scale kinematics (for motions assumed to be 100 to 200 km/s, see [39]). In other words, the turbulent gas motions observed by GIRAFFE give a good approximation of the dynamical state of both gas and star-forming regions, the latter dominating the morphological disturbances of distant galaxies. This is a good start to identify the physical mechanism(s), which reproduce the large variety of both peculiar morphologies and anomalous kinematics of spiral' progenitors.

[6] It is assumed here that local spiral galaxies are fully dominated by their rotation, but see [40] for a detailed analysis of kinematics of local galaxies.

### *Gas Fraction and Metal Content*

Conversely to local galaxies for which the gas content can be easily estimated from HI measurements [41], at z=0.65, most galaxies are out of the reach of present-day radio telescopes. On the other hand one may rely on the existing correlation between surface densities of the gas and of the star formation rate, known as the Kennicutt-Schmidt relation [42]. It is natural to expect a correlation between star formation efficiency and gas content, although the tightness of the observed relation from normal spirals to powerful ULIRGs is quite impressive as it linked very different processes generating star formation. A correlation between star formation and molecular gas is also observed [43], which is easier to understand from the theory of star formation. Gas content can be estimated from the inversion of the Kennicutt-Schmidt relation [39, 44], whose slope may slightly vary depending on the inclusion of very powerful ULIRGs. The uncertainties related to the exact slope of the KS relation are certainly marginal because the distribution of SFR surface densities is concentrated within a quite small range, $\Sigma$(SFR)=0.01-0.1 $M_O$/yr/kpc$^2$ (see Figure 5, bottom-right). Uncertainties in estimating the gas mass are probably dominated by those on the determination of the gas radii. Several distant galaxies are compact from their stellar continuum, although they almost all show relatively large extent of their ionised gas emission, which can be only detected by IFU techniques.

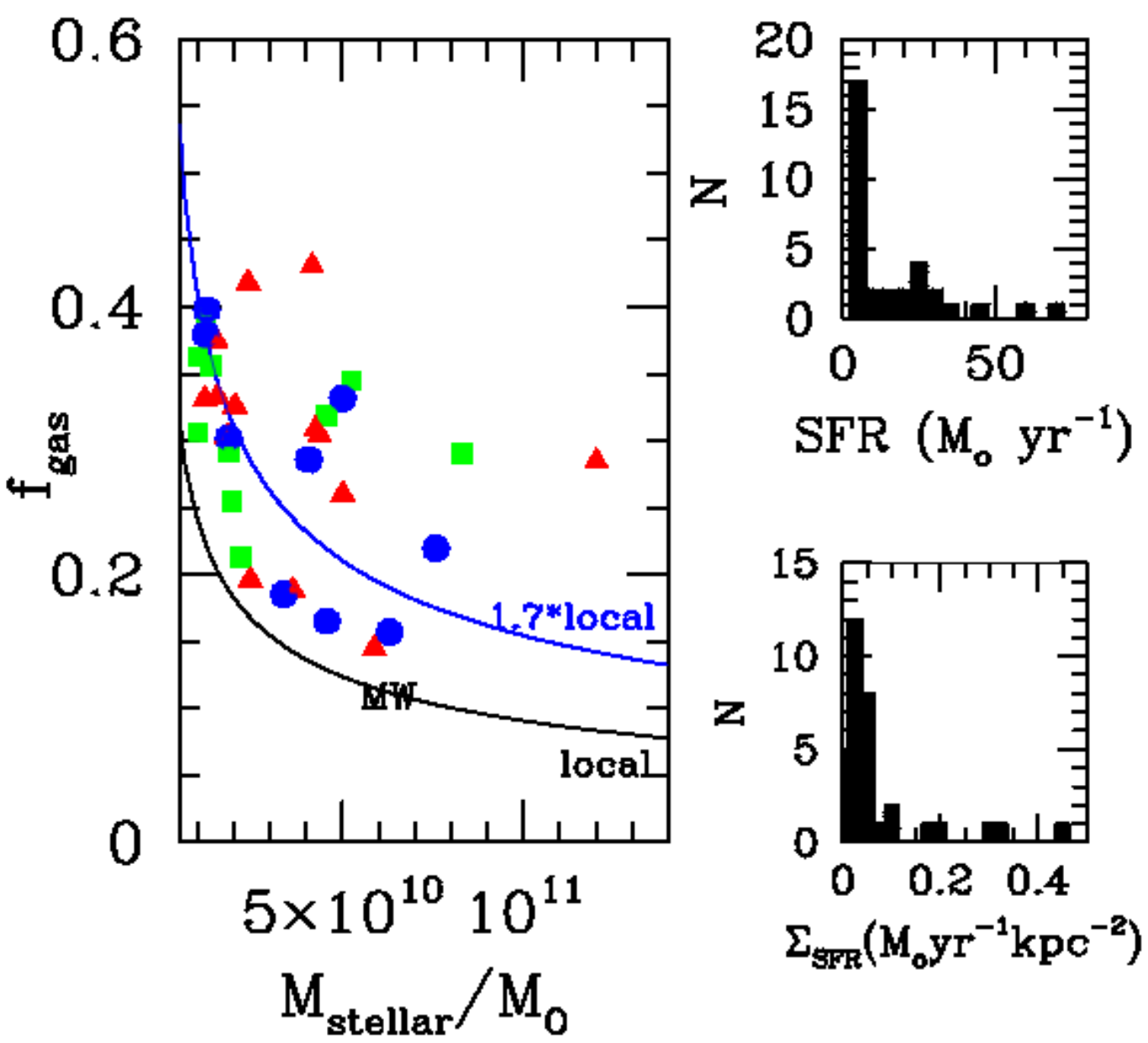

**FIGURE 5.** Gas fraction versus stellar mass for 33 z=0.4-0.75 galaxies observed in the GOODS-S field with symbols representing their kinematics as shown in Figure 4. The black curve represents the mean value for local galaxies from [41] and the location of the Milky Way is also shown. All distant galaxies lie above the local relation and assuming a similar relation between stellar mass and gas fraction implies that distant galaxies have gas fraction 1.7 times the average values at z=0. On the left the two histograms show the distribution of SFR and SFR surface density of the sample.

Figure 5 shows a median value of 31% for the gas fraction in galaxies at z=0.65. Because those galaxies (with emission lines) represent the majority of present-day spirals' progenitors, we may compare this value to that from local spirals, that is, in a similar mass range, 1.7 times smaller. Considering a sample of 65 similar galaxies at z=0.4-0.75, [45, 46] found that the Oxygen content of the gaseous phases is two times smaller at z=0.65 than at z=0. At the average stellar mass of the distant samples (3.5 $10^{10}$ $M_O$), [46] found an evolution of the O/H abundance of -0.3dex. Assuming that the average gas fraction of local galaxies is 15% (see Figure 5), one may deduce a fraction of 35-40% for z=0.65 galaxies, on the basis of a close box model (see [45, 46]). This provides an independent support to the estimate from [44] based on the inversion of the KS relation.

## Lessons from the Tully Fisher Relation and from Disk Dynamics

Mass and velocity of disk-dominated galaxies define a very tight relationship. When including all the baryonic mass, this relation follows $M_{baryonic}$=61(+60,-30) X $V_{flat}^4$ over almost five mass decades, from (gas-rich) dwarves to massive spirals [47]. Well-resolved rotational curves of nearby galaxies reveal the predominance of flat velocities in their outskirts. Interestingly most of the uncertainties in the above relationship are related to the estimate of the stellar mass budget. For local and distant galaxies we observe only the brightest stars and not the main sequence stars, which dominates the stellar mass. There are still efforts to be done to provide a better accuracy on $M_{stellar}$ [48], including using information's from the star formation rate and from a better estimate of the red stellar component based on Lick indices. Conversely the gaseous mass, except possibly that from the molecular gas, is much better estimated from HI measurements. The tightness of the "universal" baryonic Tully Fisher and the accuracy of velocity measurements allow [49] predicting the $M_{stellar}/L_K$ behaviour with colour, which is found to be consistent with estimates from [50].

It is not possible to derive accurate rotational curves for most of the distant galaxies because of the coarse spatial resolution. The problem is even more complex because of the predominance of anomalous kinematics. However [51] were able to derive $V_{flat}$ values with a reasonable accuracy, i.e. a median uncertainty of 0.12dex. This has been done by a series of Monte Carlo simulations comparing models from various rotational curves (see Figure 4) and measurements that essentially correct for the limited surface sampled by the IFU. This has been calibrated using data from local galaxies and velocities for anomalous galaxies have been derived assuming that the largest velocity gradient is indeed due to rotation. The resulting stellar-mass Tully Fisher found by IFU measurements [27, 51] show a much larger scatter (0.63dex) than that from local spirals (0.12dex), similarly to what is found from slit measurements [52]. It has been convincingly shown that a tight Tully Fisher relation still hold for rotational disks and then that all the scatter is related to galaxies with anomalous kinematics, principally those with complex kinematics [27, 51].

Adding the gas mass for the sample of 33 z=0.65 galaxies, [44] have been able to derive, for the first time, the evolution of the baryonic Tully Fisher relation. Although the result is based on only 7 distant rotating disks, they fall exactly onto the local relation. If confirmed, this result is quite fundamental to link local disks to their progenitors, 6 Giga-years ago. Rotating disks at z=0.65 could have already assembled all their baryonic mass[7] (gas and stars), and may have evolved into the present-day spirals simply by transforming two-thirds of their gaseous content into stars during the 6 Giga-years elapsed time. Evolution of fundamental relationships involving mass (e.g. Tully Fisher & mass-metallicity) is consistent with a close box model for a significant part of the present-day spirals 'progenitors. This may be extended to progenitors with perturbed kinematics, as those lie on both sides of the local baryonic Tully Fisher relation (see [44] and their Figure 5). Such a result does not exclude external sources of gas but it simply does not require them. It means that the close-box model works quite well for linking the two samples at z=0.65 and at z=0: however [46] demonstrated that this is no more the case when relating z > 2 galaxies to local galaxies, i.e. there should be other sources of gas between z> 2 and z=0.65 or z=0.

Having identified the probable mass evolution of spirals in the Tully Fisher relationship –just gas transformed into stars- I consider again the origin of the dramatic redshift increase of the scatter of the Tully Fisher relation. From the above, it is certainly not due to variation or problem with the mass estimates. Then it should be caused by strong and fundamental changes of both gas kinematics and galaxy morphologies. Another change is that of the gravitational support to the gaseous disk, which may be evaluated by the ratio V/σ. Giant thin disks have σ in the range of 10-30km/s and then V/σ from 8 to 20. For distant rotating disks and galaxies with perturbed rotation [53] found median values of 4 and 3 for V/σ, respectively. All distant galaxies present large values of σ, generally above 50km/s (see also [54]). It leads to two interesting consequences:

- Rotational gaseous disks at z=0.65 have a rotational support similar to that of the thick disk of the Milky Way (V/σ =3.9), although they appear to be the most relaxed galaxies, 6 Giga-years ago;
- A considerable fraction of the scatter in the Tully Fisher relation is removed by assuming a transfer from bulk motion to random motions (see [55] and references therein) as it can be expected in major mergers from their hydro-dynamical simulations.

---

[7] Interestingly the distant rotational disks indicate a zero-point in the stellar-mass Tully Fisher that is slightly different from that of local spirals, implying a smaller stellar mass by a factor 1.6 to 2.1 for a given velocity [51]. However the slope of the stellar mass Tully Fisher is not established on similar robust grounds than that of the baryonic Tully Fisher, and the previous factor could be well affected by small changes of the corresponding slope.

In the following I will consider the different possible mechanisms that can explain the evolutionary changes of massive galaxies.

# GAS-RICH GALAXIES MERGERS COULD FORM LARGE DISKS

## Why the major merger hypothesis is unavoidable in linking z=0.65 to z=0 galaxies?

We have to link distant galaxies with peculiar morphologies and anomalous kinematics with the numerous present-day spiral galaxies (see Figure 3). As I have shown above, such an exercise is far much easier at z=0.65 than at larger redshift, for which the existing observations are difficult to be compared to those of local galaxies. At first order, most of the mass changes since the last 6 Gyrs is due to the gas-to-stars transformation. The strong link between morphological and kinematical disturbances let few physical mechanisms able to account for them. Physical processes related to star formation have been advocated to explain the large dispersions observed in z=2 galaxies [56]. However at z=0.65 no significant outflows have been detected by [39] from the comparison of emission lines and absorption lines systems. This is not surprising because all the galaxies considered here have large baryonic masses, in excess of $10^{10}M_O$, and not very large star formation rates: as such the energy developed by massive stars is always negligible when compared to that from gravitational motions. A considerable effort has been made to explain the unusual properties of very distant galaxies at z=2. Cosmological simulations predict large reservoirs of cold gas in the very distant Universe, which prompted [57] to propose cold and clumpy gas flows to explain the high star formation in z=2 galaxies. Given the uncertainties in linking z > 2 galaxies to local ones, it is still unclear whether this scenario reproduce the progenitors of spirals galaxies, or alternatively of more early-type galaxies. All in all, no such large cold gas reservoirs are predicted at z=0.65 and as such this scenario has never been considered for such redshifts.

The coeval evolution of star formation, stellar mass assembly, fraction of peculiar galaxies, interacting pair statistics and gas content have lead [24] to propose the rebuilding disk scenario. To reproduce the observed evolutions requires that 50 to 75% of present-day spirals have been reprocessed through mergers. In parallel a considerable effort has been made in simulating mergers using appropriate hydro-dynamical simulations [58, 59, 60, 61]. Given the fact that distant galaxies have large gas content, all the studies agree on the fact that gas-rich mergers may produce a new disk in their remnant phase, and that such a new disk is mainly supported by the orbital angular momentum provided by the collision. Having in hand the observations, and all the details they brought, as well as the theoretical frame, the rebuilding disk scenario, it has to be investigated whether or not both are matching together.

## Modelling the properties of distant galaxies

The wealth of the data accumulated by the IMAGES study allows us to generate a detailed model for each individual galaxy. This is a considerable effort given the fact that we have such observations for 100 galaxies. Up to now this has been conducted only for 5 distant galaxies that have been fully modelled on the basis of hydro-dynamical simulations of mergers ([38, 62, 63, 64, 65]; see Figures 6 and 7). Completion of such a project has the important goal to reproduce the present-day Hubble sequence on the sole basis of the observed galaxies having emitted their light 6 Gyrs ago[8]. A success of this project may possibly solve the question of the origin of the Hubble sequence. It would also generate important consequences on the properties of present-day spirals, including about the origin of their substructures such as bulge, spiral arms, bars and rings. This is challenged by the important question of the model uniqueness. In fact, the further evolution of the merger remnant is not unique and uncertainties on the modelling have to be considered. We might solve this important issue by considering the conservation of important quantities, such as the mass and the angular momentum. For example, remnants of distant galaxies have to produce present-day spirals that delineated the Tully Fisher relation and the angular momentum velocity relationship.

.

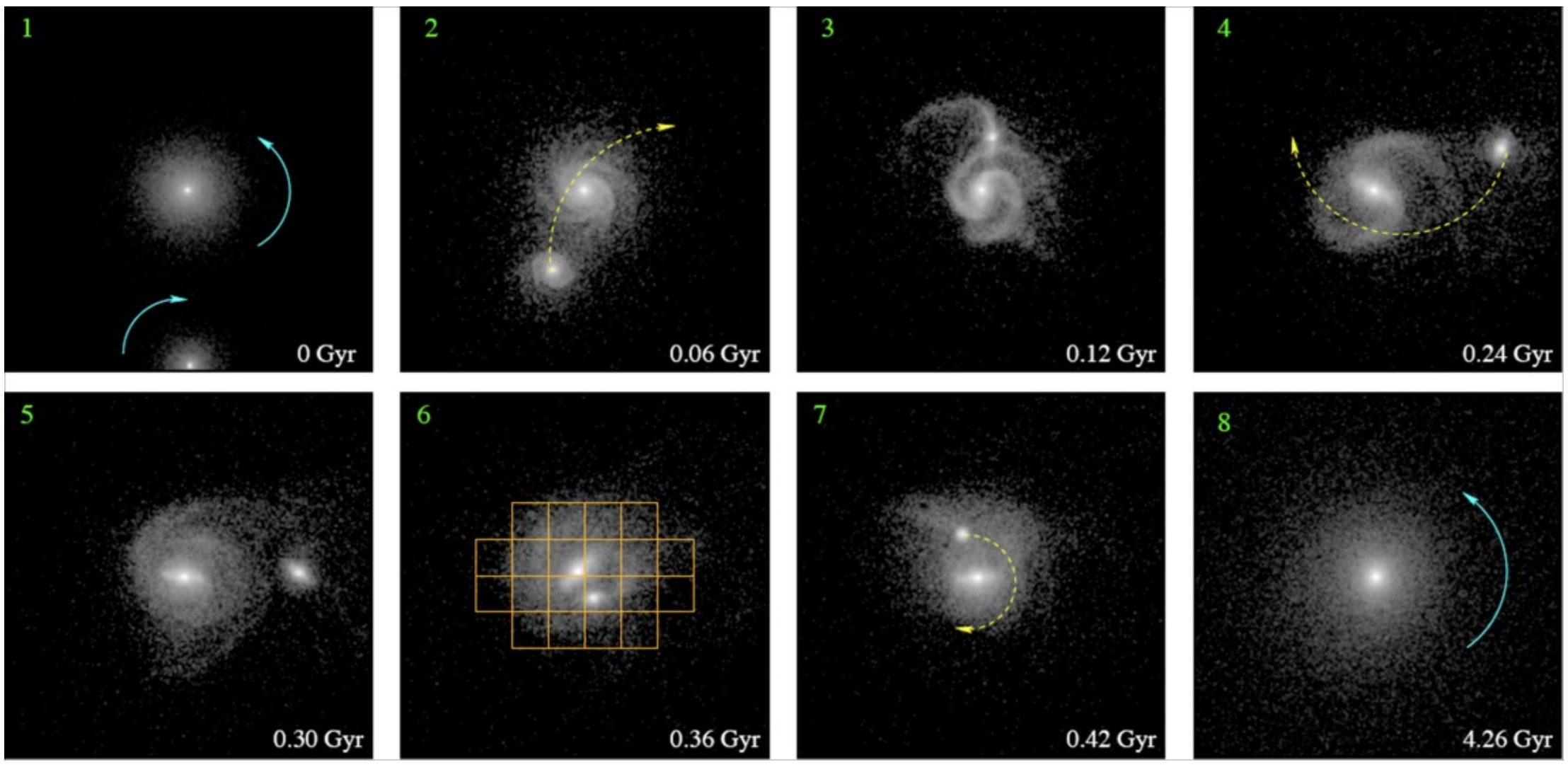

**FIGURE 6.** It shows the evolution of a 1:3 merger that reproduces both the morphology and kinematics of a distant galaxy (6th panel, [62]). The simulation has been performed using GADGET2 with stellar and gaseous masses provided by the observations. The spin of the smaller component is inverted relatively to that of the main galaxy. The interaction is responsible of the formation of the observed blue-giant bar as well as the dynamical axis that passes through the minor component. Letting the simulation evolving to z=0 (8th panel) provide a galaxy that is an S0_a: this is the object in the sample for which the simulation produces the largest B/T value because of the spin inversion.

## Major mergers can reproduce all anomalous distant galaxies

In this section I present results based on the sample of 32 galaxies that have been observed in the GOODS_S field, and including 26 galaxies showing peculiar morphologies and/or anomalous kinematics. For reasons of simplicity, [39] have assumed a limited series of simulations provided by [66], i.e. 48 different merger models. For half of the anomalous galaxies, [39] has been able to distinguish the two components by using the exquisite spatial resolution of the HST/ACS. Among these 13 anomalous galaxies, 10 have both their morphologies and kinematics well reproduced by the above simple model. Interestingly the remaining 3 galaxies are obvious cases of merger, and the failure of the model is certainly linked to its simplicity. All the 13 galaxies but one show a mass ratio larger than 1:4 implying that minor merger is not the dominant process. This is because minor mergers should have a much lower efficiency to distort galaxy morphologies and kinematics [60]. In several cases, including those having been studied at depth, there is no compelling mechanism other than merging to explain the whole properties of the galaxy (see Figure 7).

What could be the nature of the remaining anomalous galaxies showing only a single component? Detailed simulations of galaxy mergers by [59] indicate that for a significant period of 1 to 1.5 Gyr after the fusion, the remnant shows distorted morphology and kinematics. However it is uneasy to evaluate the mass ratio and the orbit of those remnants. It has been a long exercise for me (several months) to examine all these individual galaxies and to try to derive some clues of their past history. The problem is even more complex because in some cases the remnant is dusty because star formation is quite active during that phase (e.g. [59]). Indeed by using different techniques, including a full model of the light distribution of each galaxy, [39] have been able to evidence the presence of features such as bars, rings and other central features, sometimes after removing the adopted disk-plus-bulge model. This has been well illustrated by [63] for a galaxy (see Figure 8) that is probably an archetype of a disk that is rebuilt few 100 Myrs after a major fusion. The dynamical axis is 45° offset from the optical axis of the dusty disk that produces stars at a rapid rate (20 $M_O$/yr). The central features (see Figure 8 top-left panels) are helicoidally structured and are likely the response of the system to the gravitational torque after the fusion (see e.g. [60]) of galaxies with mass ratio larger than 1:3 and an inclined orbit. Accounting for the star formation in the disk, the system will evolve into a late-type spiral, with bulge to total ratio less than 0.2. On the basis of central features within the 13 anomalous galaxies with a single component, [39] have been able to well reproduce 7 of them by a major merger model. This let the major merger hypothesis as unrivalled to reproduce both kinematics and morphology of distant galaxies.

---

[8] Of course a limitation to such an ambitious project is due to the fact that it neglects possible further interactions for the remnant with other galaxies but this effect should be small as discussed in the Appendix of [26].

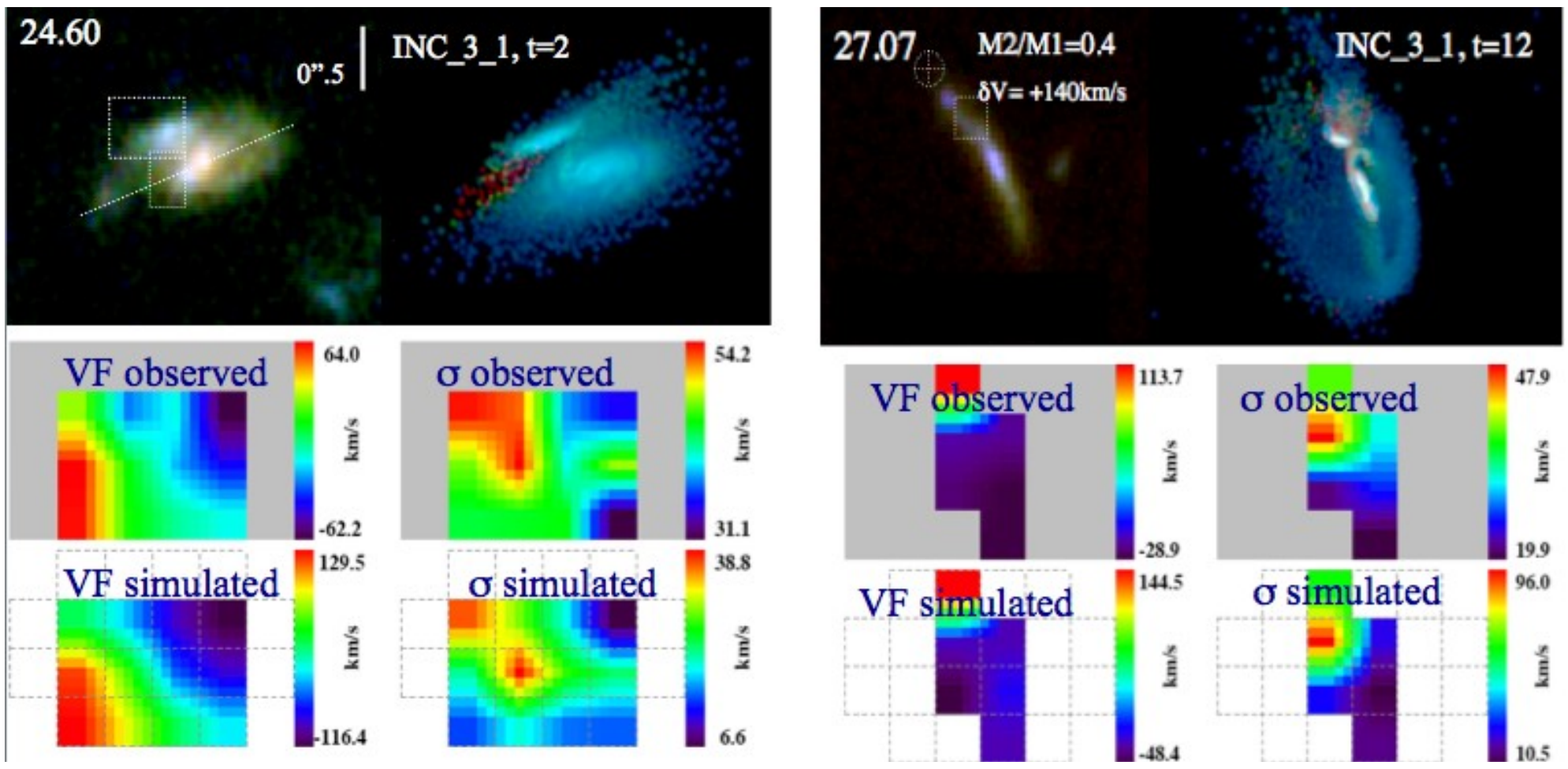

**FIGURE 7.** It shows two galaxies (J033224.60-274428.1 at z=0.5368 and J033227.07-274404.7 at z=0.7381) for which two components have been identified and which have been simulated. Images of galaxies are composite of b+v, I and z filters of HST/ACS and on the right, simulations show the gas component with the coding of the [66] model. The second galaxy on the right is a good example of an object that could not be explained by any other process than a merger, in a quite subtle way. At first glance it could resemble a giant luminous arc (e.g. [67]), but the absence of any cluster and the moderate redshift of the galaxy rules out this possibility. On the other hand it could resemble an edge on galaxy, but the velocity gradient along 90% of its length is less than 10 km/s, which again rules out this alternative. It let the single possibility that it is a giant bar inserted in a face-on spiral that has been activated by a merger for which the smaller component is identified near the dispersion peak, on the top of the galaxy.

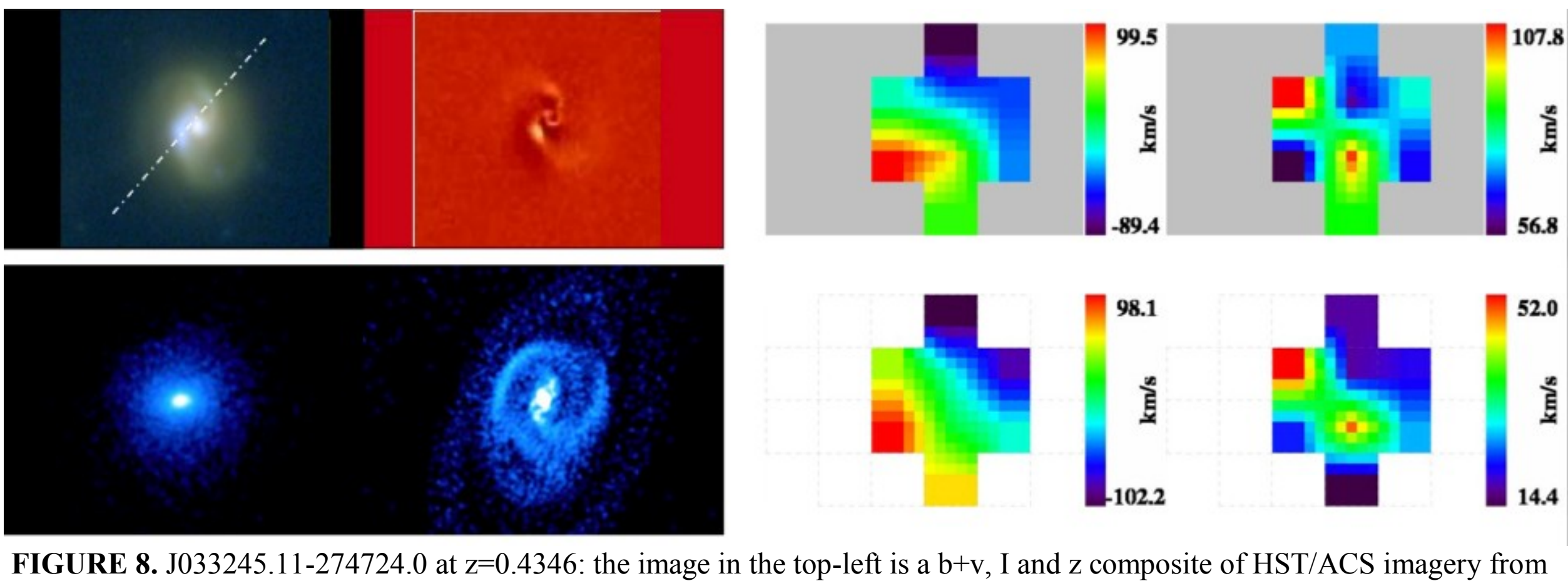

**FIGURE 8.** J033245.11-274724.0 at z=0.4346: the image in the top-left is a b+v, I and z composite of HST/ACS imagery from the UDF. The disk is much redder than the central region and because of its large dust content it can be hardly detected by shallower imagery from GOODS. The dynamical axis is indicated by the dotted-dash line and come from the observed velocity field displayed in the third top panel. The second top panel shows the residual of the galaxy after subtracting the best fit of the galaxy light from a bulge-plus-disk model. The bottom panels show modelled stars, gas distribution, velocity field and dispersion map (from left to right).

## Properties of the merging progenitors of distant galaxies

Following the assumption that most anomalous galaxies can be major mergers or their remnants, it may be of interest to investigate their progenitors properties. Figure 9 shows the observed gas fraction as well as the derived value for progenitors. To estimate the last value, I have estimated the characteristic time of star formation ($T_{SFR}$= $M_{stellar}$/SFR) for all galaxies classified in the 4 different merger phases. For each modelled galaxy, we know from the model the time it has spent during the merger. For each merger phase we can derive the median $T_{SFR}$, and reasonably assume that it is representative of each galaxy of this phase. Comparing the two times for each galaxy we can derive the fraction of gas that has been transformed into stars during the event. It results that most progenitors have gas fraction well above 50%, which is the condition to reform significant disks after the merger [58, 59, 60, 61]. The progenitors of these assumed mergers were individual galaxies at the time they were pairs, such a look-back time can be evaluated for each observed object in each phase. For example the 8 galaxies which are modelled near fusion would have their progenitors in pairs approximately 1.8 Gyrs before z=0.65, which corresponds to z=1. The median stellar mass of their progenitors would be 5 $10^9$ $M_O$ and their gas fraction 80%. Gas fraction of similar galaxies (same stellar mass) at z=0 average to 30%, and it is not surprising that the gas content of similar galaxies is more than twice at z=1. The numbers displayed in that section slightly differ from those of [39], because they account for a more realistic distribution of the characteristic merger times (see caption in Figure 9). Nevertheless they are even more compelling to support that the disk-rebuilding scenario is a serious alternative to form the present-day massive disks.

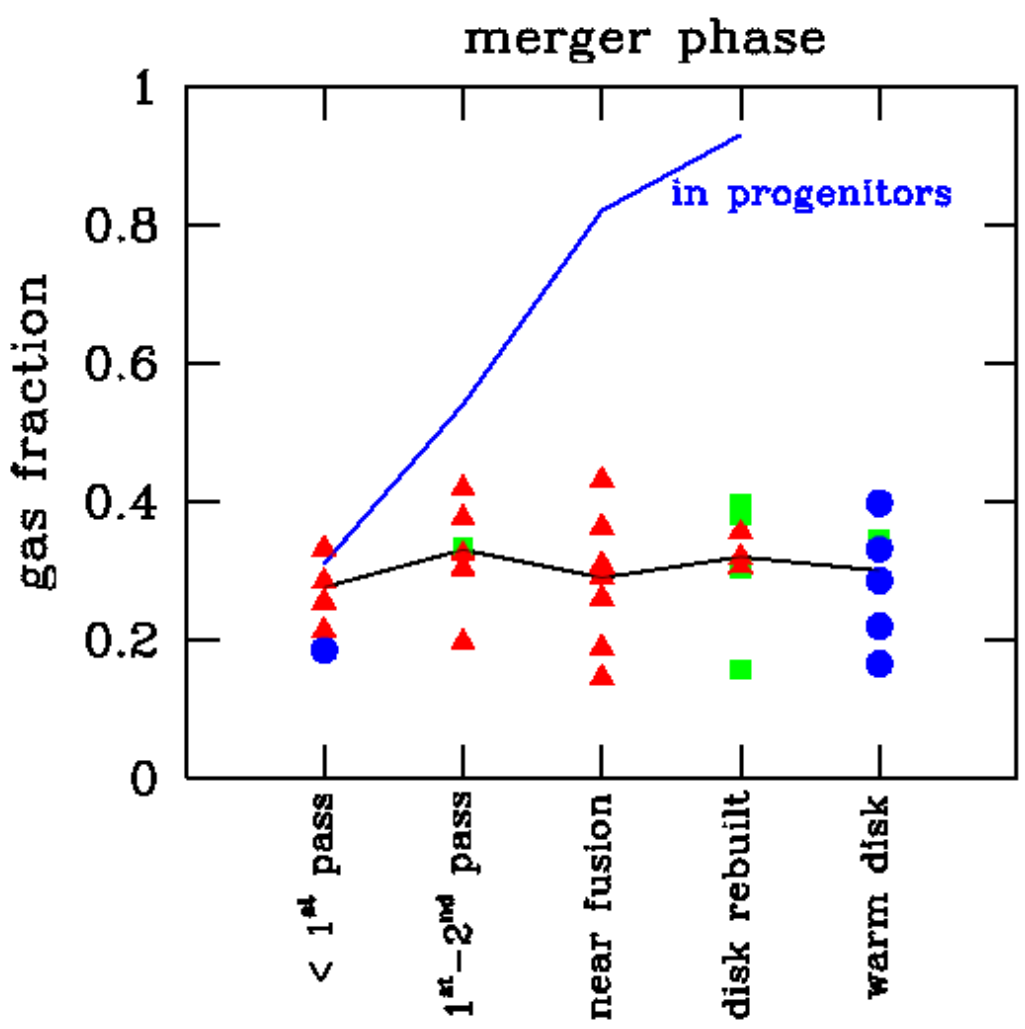

**FIGURE 9.** Observed gas fraction in distant galaxies from IMAGES (symbols are the same than in Figure 5). This Figure is an up-dated version of Figure 7 of [39] for which I have considered that the characteristic dynamical time scale with mass at the power 1/4 rather than 1, in agreement with [66]. The blue line indicates the gas fraction in the progenitors assuming that some gas has been transformed to stars during the merger (see text).

## The fraction of stars that are formed during major mergers since the last 8 Giga-years

A kind of general consensus seems to be emerging in establishing a small contribution of mergers to star formation (about 10%, see [68, 69, 70] and references therein). In fact I think there is still a lot of misunderstanding in estimating such a contribution. For example, [68, 69] have considered samples including a large number of galaxies and have assumed that in most irregular galaxies, mergers do not activate star formation. However, in absence of kinematic data, this conclusion might be premature: [31] found a very good correlation between morphological irregularities (stars) and anomalous kinematics (ionised gas). Either the IMAGES sample is too limited or biased, or the conclusion of [68, 69] requires an unknown process to distort both morphology and kinematics. Another way to minimise the impact of merging in star formation could be to limit the merger-like activity to the only peaks of star formation that are merger induced, e.g. during the first passage and at the fusion, during periods of $10^8$ years [70]. This contrasts with simulations of star formation history during mergers by [59] and [71] showing that star formation is enhanced by pressure driven gas during 1 to 3 Gyrs, including for a significant period after the merger[9]. Taking into account these simulations and results from [70], star formation induced by mergers may well account for most of the star formation evolution since the last 8 Gyrs. A more substantial effort should be done to evaluate this, by comparing from simulations how would be the star formation in isolated galaxies assuming they wouldn't have any merger during their evolution. For example, low-surface-brightness galaxies are showing large gas fractions, while they do not reach the threshold for star formation in the Kennicutt-Schmidt relation [72]. If such galaxies were involved in an interaction or a merger, gravitational forces would easily pressure the gas and then activate the star formation.

## CONCLUSION: A NEW CHANNEL TO FORM LARGE DISKS

The formation of large galactic disks has been discussed since the early 60s with a general description within massive haloes by [73]. In the frame of the tidal torque theory [2, 3], giant disks have acquired their angular momentum at very high redshift, and are assumed to evolve in a quiescent way by a smooth accretion of gas from the intergalactic medium. In this quiescent mode, substructures in spirals are mostly formed through internal instabilities. Concerning the formation of substructures in disk, a thorough review of secular evolution has been done by [74], and a recent comparison of the different channels of disk formation can be found in [75]. However the observations have considerably improved, especially those of distant galaxies. Before the 80s, they were observed as seeing limited images on either photographic plates or CCDs. Nowadays, it is possible to derive their global physical properties from photometry at UV to mid-IR wavelengths, and moreover HST and very large telescopes allow detailing their morphologies to several hundred of parsecs and their internal kinematics to few kilo-parsecs.

Since the study by [76] and subsequent works, the role of mergers has been re-examined to be an important channel for the formation of elliptical galaxies. In the corresponding simulations, the two progenitors were gas-poor spiral galaxies. Several Giga-years ago, galaxies were much richer in gas than today, and their mergers unavoidably lead to giant disks superposed to a bulge. The disk-rebuilding scenario [24, 39] propose a new channel to form disks that would have acquired their angular momentum much more recently than in the tidal torque theory. The latter theory appears quite efficient in reproducing most of the properties of the Milky Way, although the representativeness of the Milky Way is questioned [15]. It however faces with two major difficulties: the disks formed by an early acquisition of their angular momentum appear too small when compared to observations, and it is at odd with the discovery that a large fraction of spiral' progenitors had peculiar morphologies and anomalous kinematics, 6 Giga-years ago.

The new proposed channel of disk formation accounts by principle to all the evolutionary features that have been observed when comparing distant to nearby spirals [24]. It might solve the angular momentum crisis as illustrated by [19], although this needs to be confirmed on actual data. In such a model, a galaxy like M31 could have been formed after a gas-rich merger at z~ 1 between two galaxies having each a stellar mass of 8 $10^9 M_O$, i.e. twice the value for M33. Such progenitors have to be gas-rich to rebuild a significant disk after the collision, and the required gas-richness is found to be consistent with high redshift observations [39]. Ages of stars in present-day disks have to be investigated to evaluate whether these disks can be enough young to have been formed more recently than at z=2 to 3. This can be done in a robust way only on galactic disks in our neighbourhood, as is has been shown by [15] for M31 with ages up to 8 Giga-years. For more distant galaxies, derivations of ages from spectral energy distributions cannot discriminating in a robust way stellar ages from 8 to 10-11 Giga-years, i.e. stars formed from z=1 to z=2-3. If true, the spiral-rebuilding scenario could imply that the whole Hubble sequence is mainly a consequence of galaxy mergers, those being predicted by the hierarchical formation of galaxies. To prove or disprove it, one has to compare it to the merger rate derived from the halo-occupation method [70] and to show how the angular momentum of distant galaxies can be linked to that of present-day spirals. Simulations have also to reproduce the present-day thin disks from the anomalous galaxies and thick disks that dominate the Universe, 6 Giga-years ago.

[9] To illustrate this point, let us compare it to a medical survey evaluating the impact of the H1N1 flux and which would account for only the sick persons having high fever, discarding all the persons who are affected at other stages of the illness.

## ACKNOWLEDGMENTS

This review has been written on the basis of the work of numerous collaborators, mostly from the IMAGES team. Such a study would not have been achieved without the important contributions of Yanchun Liang, Yanbin Yang, Xianzhong Zheng, Benoit Neichel, Lia Athanassoula, Sebastien Peirani, Myriam Rodrigues and Rodney Delgado-Serrano.